%% file: main.tex
\documentclass[journal=jacsat,manuscript=article]{achemso}

\usepackage{chemformula} 
\usepackage[T1]{fontenc} 
\usepackage{epstopdf}
\usepackage{comment}
\usepackage{hyperref}

\author{Ian Addison-Smith}
\affiliation[USM]
{Department of Mechanical Engineering, Universidad T\'ecnica Federico Santa Mar\'ia, Valpara\'iso, Chile.}
\author{Horacio V. Guzm\'an}
\affiliation[JSI]
{Department of Theoretical Physics, Jožef Stefan Institute, Jamova 39, 1000 Ljubljana, Slovenia.}
\alsoaffiliation[UAM]
{Departamento de F\'isica Te\'orica de la Materia Condensada, Universidad Aut\'onoma de Madrid, E-28049 Madrid, Spain.}
\author{Christopher D. Cooper}
\affiliation[USM]
{Department of Mechanical Engineering, Universidad T\'ecnica Federico Santa Mar\'ia, Valpara\'iso, Chile}
\alsoaffiliation[CCTVal]
{Centro Cient\'ifico Tecnol\'ogico de Valpara\'iso, Valpara\'iso, Chile}
\email{christopher.cooper@usm.cl}

\title[]
  {Accurate boundary-integral formulations for the calculation of electrostatic forces with an implicit-solvent model}

\abbreviations{IR,NMR,UV}

\begin{document}

\begin{tocentry}
\includegraphics[width=1\textwidth]{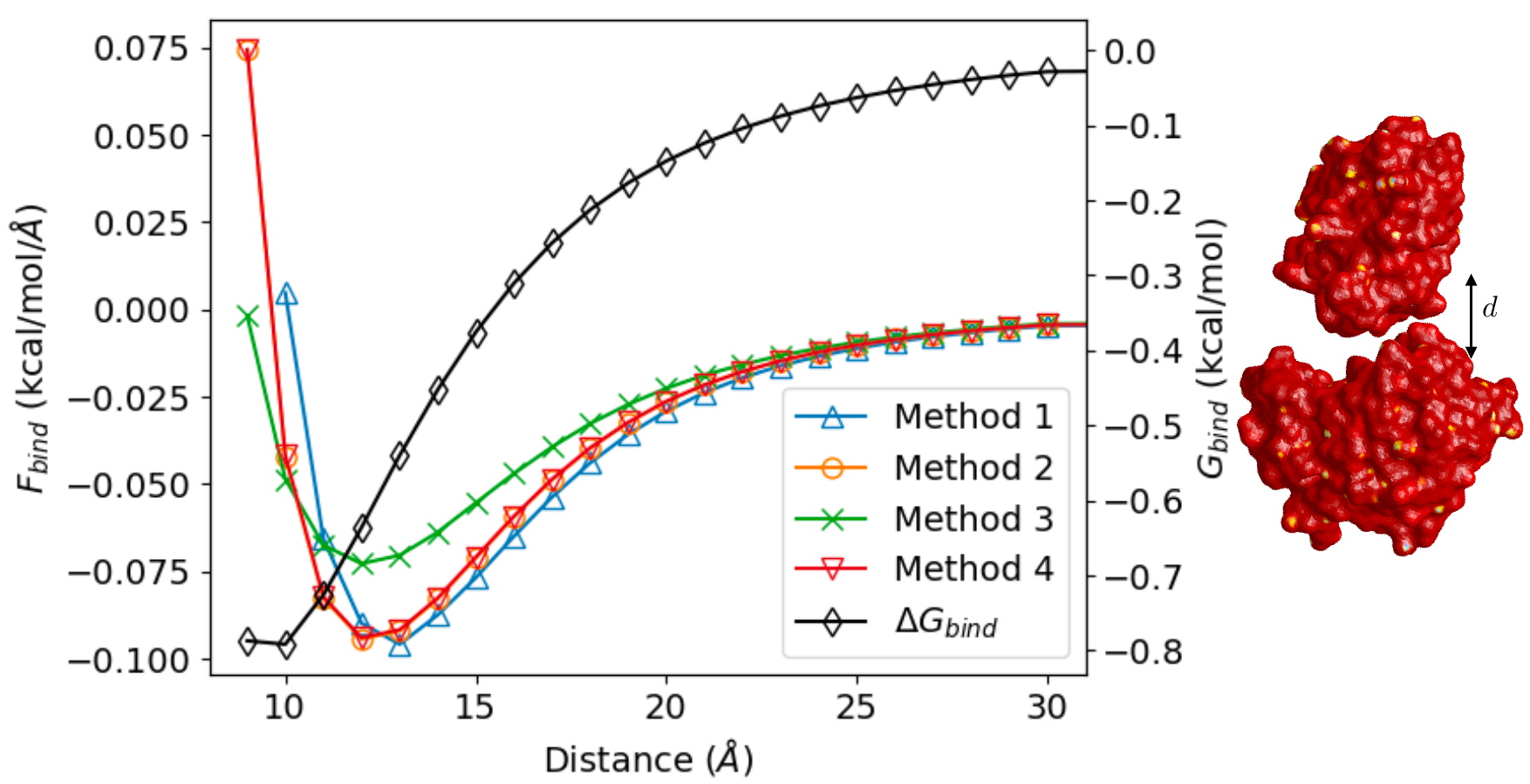}
\end{tocentry}
\newpage
\section{Abstract}
\input{Abstract}

\newpage
\section{Introduction}
\input{intro}

\section{Methods}
\input{methods}

\section{Results and discussion}
\input{results}


\section{Conclusions}
\input{conclusions}

\section{Acknowledgements}
Financial support for this project has been provided by Universidad T\'ecnica Federico Santa Mar\'ia through project PI-LIR-2020-10. C.D.C. acknowledges the support from CCTVal through ANID PIA/APOYO AFB220004 . H.V.G thanks the financial support by the Slovenian Research Agency (Funding No. P1-0055) and the financial support  of the Community of Madrid and the European Union through the European Regional Development Fund (ERDF), financed as part of the Union response to Covid-19 pandemic.

\section{Conflicts of Interest}
Authors declare no conflict of interest related to the material. 

\section{Supplementary information}
All the code and data required to reproduce the results of this work can be found in the repository at \url{https://github.com/bem4solvation/paper_PBforces}.

\bibliography{main}






\end{document}

%% file: abstract.tex
An accurate force calculation with the Poisson-Boltzmann equation is challenging, as it requires the electric field on the molecular surface. Here, we present a calculation of the electric field on the solute-solvent interface that is exact for piece-wise linear variations of the potential and analyze four different alternatives to compute the force using a boundary element method. We performed a verification exercise for two cases: the isolated and two interacting molecules. Our results suggest that the boundary element method outperforms the finite difference method, as the latter needs a much finer mesh than in solvation energy calculations to get acceptable accuracy in the force, whereas the same surface mesh than a standard energy calculation is appropriate for the boundary element method. Among the four evaluated alternatives of force calculation, we saw that the most accurate one is based on the Maxwell stress tensor. However, for a realistic application, like the barnase-barstar complex, the approach based on variations of the energy functional, which is less accurate, gives equivalent results. This analysis is useful towards using the Poisson-Boltzmann equation for force calculations in applications where high accuracy is key, for example, to feed molecular dynamics models or to enable the study of the interaction between large molecular structures, like viruses adsorbed onto substrates.

%% file: intro.tex
Implicit-solvent models consider a dissolved molecule as a cavity inside an infinite dielectric medium, averaging out the discrete degrees of freedom of the solvent \cite{RouxSimonson1999,decherchi2015implicit}, which yields an efficient way to compute mean-field potentials and free energies. A popular version of these models uses the Poisson-Boltzmann equation to represent the electrostatic potential in an ionic solvent \cite{Baker2004}. Numerical solutions of this equation are implemented in a variety of solvers that use finite difference,\cite{baker2001electrostatics,jurrus2018improvements,gilson1988calculating} finite element,\cite{baker2001electrostatics,bond2010first} or boundary element (\texttt{BEM}) methods.\cite{boschitsch2002fast,Lu2005a,geng2013treecode,CooperBardhanBarba2014}

Most applications of the Poisson-Boltzmann model apply it to compute the mean field electrostatic potential  and polar component of the solvation energy, however, it can also compute the electrostatic force \cite{AlexovDelphiForce, AFMPB, baker2001electrostatics, jurrus2018improvements,jha2022computation}. This force is useful to study the interaction between multiple bodies \cite{cooper2022quantitative}, which can be fed into molecular dynamics codes ({\it i.e.} for docking \cite{AlexovDelphiForce}). 

There are three ways to compute the force with the Poisson-Boltzmann equation: starting from the variation of the energy functional \cite{Gilson1993, Im1998, Davis1990}, using the Maxwell stress tensor\cite{Lu2005a,Lu2005b,bordner2003} or calculating the variation of the solvation energy numerically \cite{Davis1990}. Regardless of the method of choice, this calculation is challenging as it involves either {\it (i)} the subtraction of two large numbers \cite{Gilson1993}, {\it (ii)} calculating hypersingular integrals \cite{Lu2005b}, or {\it(iii)} numerical differentiation across the molecular surface \cite{LuETal2009}. It is also model-dependent, as there are differences if the dielectric interface is sharp or continuous \cite{Xiao2013}. Moreover, if the Poisson-Boltzmann equation is being solved with a finite difference method, the electric field on the molecular surface is computed with a mollified interface\cite{AlexovDelphiForce, jurrus2018improvements} or approximated with least squares,\cite{boschitsch2015adaptive} which may introduce a diffusive effect to the solution. The boundary element method offers a more accurate description of the molecular surface, however, current implementations do not overcome the limitations described earlier \cite{CiCP-3-973}. Alternatively, we can reformulate the expressions resulting from taking the variation of the energy functional and the Maxwell stress tensor in terms of an apparent surface charge~\cite{zauhar1991,cortis1997,bordner2003}. Also, analytical calculations of the force are possible when using the conductor-like screening model (\texttt{COSMO}) type models.\cite{jha2022computation}

The goal of this work is two-fold. First, we present a new formulation to compute the electric field across the boundary that is exact for piece-wise linear boundary elements. This allows us to compute the force without adding numerical approximations on top of standard electrostatic potential calculations. Second, we perform a thorough assessment of the accuracy of the force computed with different methods, implemented in the Poisson-Boltzmann \& Jupyter (\texttt{PBJ}) code \cite{SearchCooperWout2022}.

In the next section we present the implicit solvent model, and how the Poisson-Boltzmann equation is formulated with a boundary integral approach. This section also gives details on the calculation of the energy and force in a Poisson-Boltzmann continuum. In the Results and Discussion section we show the accuracy of the different methods for the force calculation, in settings with isolated and interacting molecules. The final section presents conclusions and outlook for future work.

%% file: methods.tex
\subsection{The Poisson-Boltzmann equation with a boundary integral formulation}

In the context of molecular solvation, the Poisson-Boltzmann model considers the solute as a low-dielectric cavity immersed in an infinite continuum domain. Following Fig. \ref{fig:molecule}, the {\it solute} region ($\Omega_1$) has point sources to represent the partial charges ($q_k$), and is contained inside the molecular surface ($\Gamma$). There are several possible definitions of $\Gamma$, such as the solvent-accessible, solvent-excluded, van der Waals, and Gaussian surfaces. We chose the solvent-excluded surface (SES),\cite{Connolly83} which is the result of tracking the contact points between the solute and a spherical probe that is rolled around it. On the other hand, the external region corresponds of an ionic solvent (usually, water with salt). The free ions in the solvent have an effect on the electric field, and if they are considered as point charges that arrange according to Boltzmann statistics, continuum electrostatic theory leads to the (linearized) Poisson-Boltzmann equation. We can express this as the following system of partial differential equations
\begin{align} \label{eq:pde}
\nabla^2\phi_1 = \frac{1}{\epsilon_1} \sum_{k=1}^{N_q} q_k\delta(\mathbf{x}_k) \quad &\mathbf{x}\in\Omega_1 \nonumber\\
\left(\nabla^2-\kappa^2\right)\phi_2 = 0 \quad &\mathbf{x}\in\Omega_2 \nonumber\\
\phi_1 = \phi_2; \quad \epsilon_1\frac{\partial\phi_1}{\partial\mathbf{n}} = \epsilon_2\frac{\partial\phi_2}{\partial\mathbf{n}} \quad &\mathbf{x}\in \Gamma.
\end{align}
where $\phi$ is the electric potential, $\kappa$ is the inverse of the Debye length, $\delta(\mathbf{x}_k)$ is the Dirac delta function at $\mathbf{x}_k$ and $\mathbf{n}$ a unit vector that is normal to $\Gamma$.

\begin{figure}
    \centering
    \includegraphics{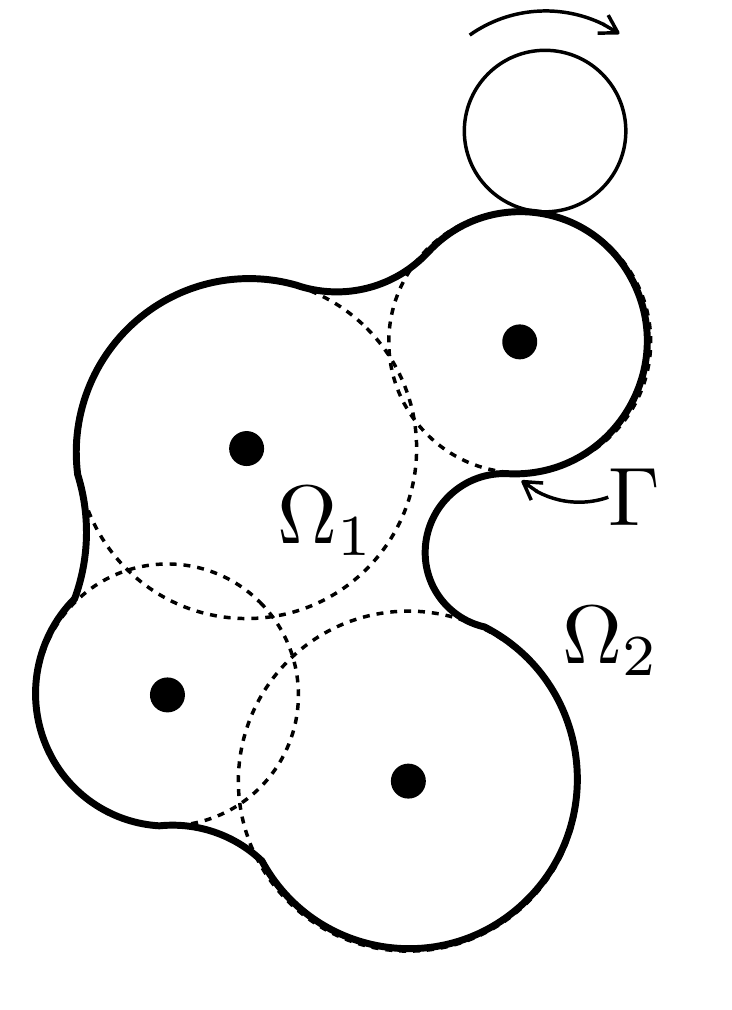}
    \caption{Representation of a solute in a continuum model.}
    \label{fig:molecule}
\end{figure}
    
\subsubsection{The boundary integral formulation}

A common approach is to formulate Eq. \eqref{eq:pde} as an integral over $\Gamma$. Applying Green's second identity to Eq. \eqref{eq:pde}, we arrive at
\begin{align}\label{eq:bie}
\phi_{1}(\mathbf{x}) &= -K_{\Gamma,L}^{\mathbf{x}}\left(\phi_{1, \Gamma}\right) + V_{\Gamma,L}^{\mathbf{x}}\left(\dfrac{\partial}{\partial \mathbf{n}} \phi_{1, \Gamma}\right) +\dfrac{1}{\epsilon_{1}} \sum_{k=1}^{N_{q}} \dfrac{q_{k}}{4\pi |\mathbf{x} - \mathbf{x}_{k}|} \quad \mathbf{x}\in\Omega_1 \nonumber\\
\phi_{2}(\mathbf{x}) &= K_{\Gamma,Y}^{\mathbf{x}}(\phi_{2, \Gamma}) - V_{\Gamma_,Y}^{\mathbf{x}}\left(\dfrac{\partial}{\partial \mathbf{n}} \phi_{2, \Gamma}\right) \quad \mathbf{x}\in\Omega_2 \end{align}
where $\mathbf{x}\in\Omega_1\cup\Omega_2\textbackslash \Gamma$. Also, 
\begin{align}
V_{\Gamma}^{\mathbf{x}}\left(\varphi\right) &= \oint_{\Gamma} G(\mathbf{x},\mathbf{x}')\varphi(\mathbf{x}')\rm{d}\mathbf{x}'\nonumber\\
K_{\Gamma}^{\mathbf{x}}\left(\varphi\right) &= \oint_{\Gamma} \frac{\partial G}{\partial\mathbf{n}}(\mathbf{x},\mathbf{x}')\varphi(\mathbf{x}')\rm{d}\mathbf{x}'
\end{align}
are known as the single- and double- layer potentials, and 
\begin{align}
G_L(\mathbf{x},\mathbf{x}') &= \frac{1}{4\pi|\mathbf{x}-\mathbf{x}'|} \nonumber\\
G_Y(\mathbf{x},\mathbf{x}') &= \frac{e^{-\kappa|\mathbf{x}-\mathbf{x}'|}}{4\pi|\mathbf{x}-\mathbf{x}'|} 
\end{align}
are the free-space Green's function of the Laplace and Yukawa (Poisson-Boltzmann) potentials. 

Combinations of the expressions in Eq. \eqref{eq:bie} yield different boundary integral formulations,\cite{SearchCooperWout2022} that vary in complexity and the conditioning of the resulting matrix. Here we use the simplest form, termed {\it direct formulation},\cite{YoonLenhoff1991} which is implemented in the \texttt{PBJ} code\cite{SearchCooperWout2022}.
The direct formulation simply takes the limit of the expressions in Eq. \eqref{eq:bie} as $\mathbf{x}\to\Gamma$, leaving
\begin{align}\label{eq:bie_limit}
\dfrac{\phi_{1, \Gamma}}{2} + K_{\Gamma,L}^{\Gamma}\left(\phi_{1, \Gamma}\right) - V_{\Gamma,L}^{\Gamma}\left(\dfrac{\partial}{\partial \mathbf{n}} \phi_{1, \Gamma}\right) &= \dfrac{1}{\epsilon_{1}} \sum_{k=1}^{N_{q}} \dfrac{q_{k}}{4\pi |\mathbf{x}_{\Gamma} - \mathbf{x}_{k}|} \nonumber\\
\dfrac{\phi_{1, \Gamma}}{2} - K_{\Gamma,Y}^{\Gamma}(\phi_{1, \Gamma}) + \frac{\epsilon_1}{\epsilon_2} V_{\Gamma,Y}^{\Gamma}\left(\dfrac{\partial}{\partial \mathbf{n}} \phi_{1, \Gamma}\right) &= 0
\end{align}
There are many other boundary integral formulations of this problem\cite{SearchCooperWout2022} that yield better conditioned systems than Eq. \eqref{eq:bie_limit}, for example, Juffer's\cite{juffer1991electric} and Lu's\cite{lu2007} formulations. The force calculation presented in this work is applicable to any formulation. 

\subsection{Energy in a Poisson-Boltzmann continuum}

In a continuum description, the electrostatic free energy is a function of the electrostatic potential ($\phi$), the charge distribution in the solute $(\rho_f = \sum_{j=1}^{N_q} q_j \delta(\mathbf{x}_j))$ and the concentration of free ions in the solvent ($c_j$, for species $j$). At equilibrium, $c_j$ takes the Boltzmann distribution. This transforms Gauss's law into the Poisson-Boltzmann equation, and the Gibbs free energy functional takes the form\cite{che2008electrostatic}
\begin{equation}
G = \int_{\Omega} \left \{ \rho_{f} \phi -\frac{\epsilon(\mathbf{x})}{2} \left | \nabla \phi \right |^{2}  - \beta^{-1} \sum_{j=1}^{M} c_{j}^{\infty} \left (  e^{-\beta q_{j} \phi} - 1\right) \lambda \right \} d \mathbf{x}
\label{eqn:G_NPBE}
\end{equation}
where $\beta = 1/(kT)$ is the inverse thermal energy, $c_j^{\infty}$ the bulk concentration at far away of the solute at vanishing electrostatic potential, and $\lambda$ is a unit-step function that masks out the salt-free solute region. In linear form, Eq. \eqref{eqn:G_NPBE} becomes\cite{Baker2004}
\begin{equation}
    G = \int_{\Omega} \left \{  \rho_{f} \phi  -\frac{\epsilon(\mathbf{x})}{2} \left | \nabla \phi \right |^{2}  - \frac{1}{2} \epsilon \kappa^{2} \phi^{2} \lambda \right \} d \mathbf{x}
    \label{eqn:G_LPBE}
\end{equation}

At equilibrium, the free energy $G$ reaches a minimum value \cite{che2008electrostatic}. Using the Euler-Lagrange equation, the minimum is
\begin{align}
\label{eq:Gmin}
    \frac{\partial G}{\partial \phi}-\sum_{j=1}^{n} \frac{\partial}{\partial x_{j}}\left(\frac{\partial G}{\partial \phi_{x_j}}\right) = 0 &\Rightarrow \rho_f - \epsilon \kappa^2 \phi \lambda + \nabla \cdot (\epsilon(\mathbf{x}) \nabla \phi) = 0 \nonumber \\
    &\Rightarrow  \nabla \cdot (\epsilon(\mathbf{x}) \nabla \phi) = -\rho_{f} + \epsilon \kappa^{2} \phi  \lambda
\end{align}
for $x_j$ $(j\in{1,2,3})$ a component of $\mathbf{x}$. Eq. \eqref{eq:Gmin} shows that the electrostatic potential that minimizes the energy is a solution of the Poisson-Boltzmann equation. We can use the identity $\nabla\cdot\left(\epsilon\phi\nabla\phi\right) = \phi\nabla\cdot\left(\epsilon\nabla\phi\right) + \epsilon\nabla\phi\cdot\nabla\phi$ and consider $\int_\Omega\nabla\cdot\left(\epsilon\phi\nabla\phi\right)d\Omega = 0$ (as $\phi$ goes to 0 at infinity), to rewrite Eq. \eqref{eqn:G_LPBE} as
\begin{equation}
    \begin{aligned}
    G &= \int_{\Omega} \left \{ \rho_{f} \phi + \frac{1}{2} \left (  \phi \nabla \cdot (\epsilon(x) \nabla \phi ) \right )  - \frac{1}{2} \epsilon \kappa^{2} \phi^{2} \lambda \right \} d \mathbf{x} \\
     &= \int_{\Omega} \left \{ \rho_{f} \phi  +\frac{1}{2}  \phi \left (  - \rho_{f} +  \epsilon \kappa^{2} \phi \lambda ) \right ) - \frac{1}{2} \epsilon \kappa^{2} \phi^{2} \lambda \right \} d \mathbf{x} \nonumber \\
     &= \frac{1}{2}\int_\Omega \rho_{f} \phi d\mathbf{x}   
    \end{aligned}
\end{equation}
Acknowledging the charge distribution in the solute is a set of Dirac delta functions, and that the solvation process is the difference between vacuum and solvated states, we arrive at the well known expression for solvation energy
\begin{equation} \label{eq:Gint}
  \Delta G_{solv} = \frac{1}{2}\sum_{k=1}^{N_q} q_k\phi_{reac}(\mathbf{x}_k) 
\end{equation}
where $\phi_{reac} = \phi -\phi_{coul}$ is the reaction potential at the location of the atoms $(\mathbf{x}_k)$. In the context of the boundary integral formulation, $\phi_{reac}$ can be computed by subtracting out the Coulomb contribution from the first expression in Eq. \eqref{eq:bie}, as follows
\begin{equation}\label{eq:phi_reac}
\phi_{reac}(\mathbf{x}) = -K_{\Gamma,L}^{\mathbf{x}}\left(\phi_{1, \Gamma}\right) + V_{\Gamma,L}^{\mathbf{x}}\left(\dfrac{\partial}{\partial \mathbf{n}} \phi_{1, \Gamma}\right)
\end{equation}

\subsection{Forces in a Poisson-Boltzmann continuum}

\subsubsection{Virtual displacement approach}

Force is the gradient of the energy in Eq. \eqref{eqn:G_LPBE} along a  coordinate. Then, we can use the virtual work principle to compute the force by evaluating the energy at positions displaced by a small value $h$ \cite{Davis1990}, and performing a finite-difference-type calculation as
\begin{equation}\label{eq:virtual_disp}
    F_{i}(\mathbf{x}) = - \frac{\partial G}{\partial x_i} (\mathbf{x}) \approx - \left (\frac{G(\mathbf{x}+h\mathbf{e}_i) - G(\mathbf{x} - h\mathbf{e}_i)  }{2h} \right ) 
\end{equation}
Here, we can compute any component of the force by performing the displacements in the corresponding direction ($x,y,z$). This approach is convenient because it does not involve any modification of a standard Poisson-Boltzmann solver that can compute the energy. However, accuracy becomes an issue as energy differences are usually small, and the numerical solver needs to appropriately resolve the electrostatic potential, requiring meshes that are much finer than common solvation energy calculations. On top of this, it requires multiple energy calculations, increasing calculation time.

\subsubsection{Energy functional variation approach}

Gilson {\it et al.}\cite{Gilson1993} used the virtual work principle to take variations of the energy functional in Eq. \eqref{eqn:G_LPBE} to find a force density function. This is,
%
%
\begin{equation}\label{eq:force_density}
\mathbf{f}= \rho_{f} \mathbf{E}-\frac{1}{2} \left | \mathbf{E} \right |^2 \nabla \epsilon-\frac{1}{2} \epsilon \kappa^{2} \phi^{2} \nabla \lambda
\end{equation}
which can be integrated in the volume to find the total force. We refer the reader to the work by Gilson {\it et al.}\cite{Gilson1993} for the complete derivation that leads to Eq. \eqref{eq:force_density}. 

Eq. \eqref{eq:force_density} introduces a clear distinction between three sources of force : 
\begin{itemize}
    \item Charge 
    \begin{equation}\label{eq:Fq}
    \mathbf{F}_{q} = \int_\Omega \rho_{f} \mathbf{E} d\mathbf{x},
    \end{equation}   
    due to the electric field ($\mathbf{E}$) on the charges. Similar to the electrostatic potential, $\mathbf{E}$ can be decomposed into coulombic ($\mathbf{E}_{coul}$) and reaction ($\mathbf{E}_{reac}$) components.
    \item Dielectric boundary
    \begin{equation}\label{eq:Fdb}
    \mathbf{F}_{db} = -\int_\Omega \frac{1}{2}  \left | \mathbf{E} \right |^2 \nabla \epsilon d\mathbf{x},
    \end{equation}
    from the jump in $\epsilon$ across the molecular surface.
    \item Ionic boundary (osmotic pressure)
    \begin{equation}\label{eq:Fib}
    \mathbf{F}_{ib} = -\int_\Omega \frac{1}{2} \epsilon \kappa^{2} \phi^{2} \nabla \lambda d\mathbf{x},
    \end{equation}
    which appears as the ionic concentration drops to 0 inside the solute. In Eq. \eqref{eq:Fib}, $\lambda$ is a mask function that is 0 in $\Omega_1$ and 1 in $\Omega_2$.
\end{itemize}

\subsubsection{Maxwell stress tensor approach}

Starting from the volume integral of the force density in Eq. \eqref{eq:force_density}, we can use the divergence theorem to write it in terms of a surface integral as
\begin{equation}\label{eq:maxwell_integral}
\mathbf{F} = \int_\Omega \mathbf{f} d \mathbf{x} = \int_\Omega \nabla \cdot \mathbf{P} d \mathbf{x}=\oint_\Gamma \mathbf{P} \cdot \mathbf{n} d \mathbf{x}.
\end{equation}
Here, $\mathbf{P}$ is a modified version of the Maxwell stress tensor, that includes the effect of the salt concentration. Following the details in the work by Xiao {\it et al.},\cite{Xiao2013} we obtain the following expression for the components of the stress tensor
\begin{equation}\label{eq:maxwell_tensor}
   P_{ij} = \epsilon E_{i} E_{j}-\frac{1}{2} \epsilon E_k E_k \delta_{i j}-\frac{1}{2} \epsilon \kappa^{2} \phi^{2} \lambda \delta_{i j} 
\end{equation}

Different from the energy functional approach in Eq. \eqref{eq:force_density}, the Maxwell tensor does not distinguish the different sources of force. In the last term of Eq. \eqref{eq:maxwell_tensor} we find the ionic boundary force ($\mathbf{F}_{ib}$ in Eq. \eqref{eq:Fib}), however, $\mathbf{F}_q$ and $\mathbf{F}_{db}$ are mixed in the first two terms.

\begin{figure}
    \centering
    \includegraphics[width=0.6\textwidth]{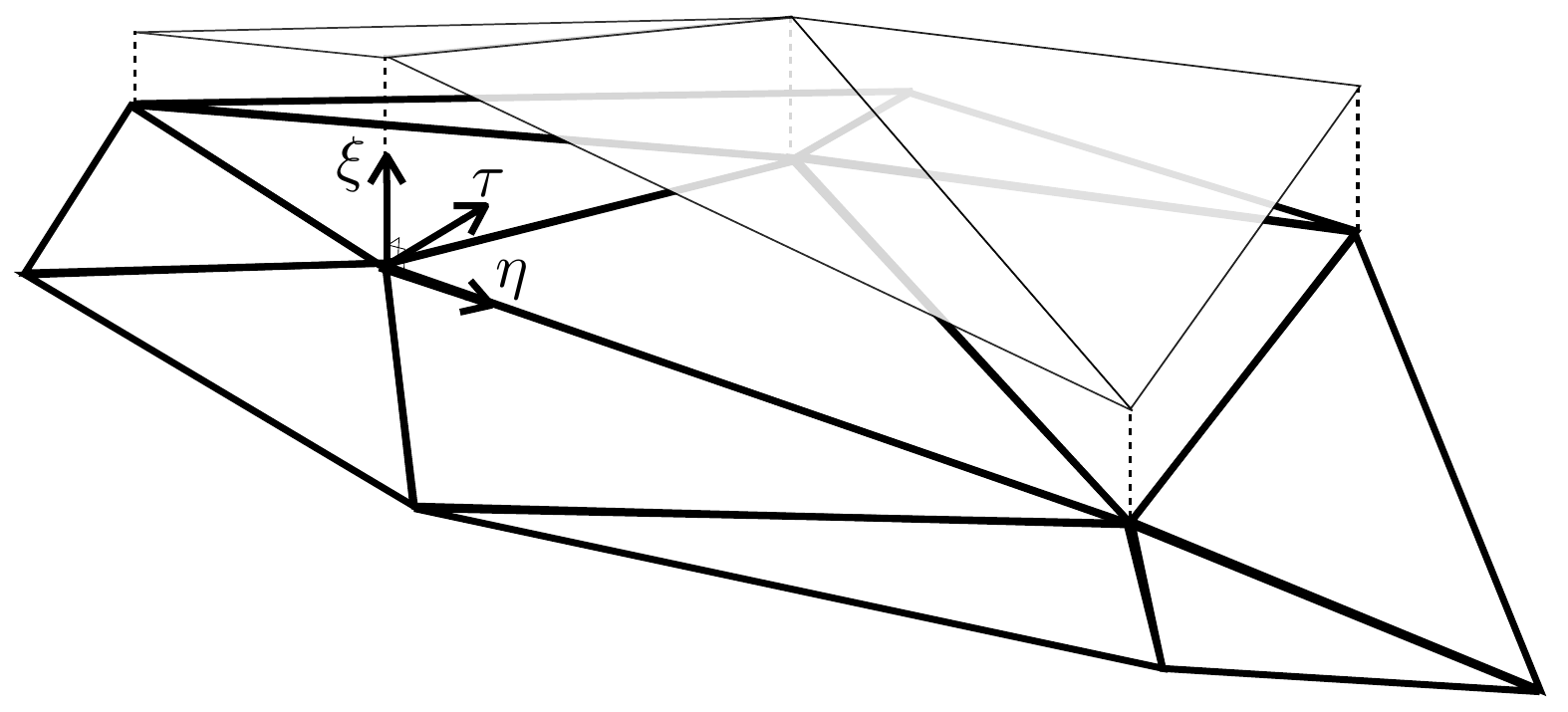}
    \caption{Local coordinate system for calculation of the force using Maxwell's stress tensor. Bottom set of triangles represents the surface mesh on the molecular surface, whereas the top set of triangles corresponds to the piece-wise linear distribution of $\phi$ and $\partial\phi/\partial\mathbf{n}$.}
    \label{fig:maxwell}
\end{figure}

The $i,j\in{1,2,3}$ indices of the Maxwell stress tensor in Eq. \eqref{eq:maxwell_tensor} usually indicate the cartesian $x$, $y$, and $z$ components. However, it can be represented in any frame of reference. Following the work by Cai and co-workers,\cite{cai2012dielectric} we use a per-element local coordinate system $\xi$, $\eta$, $\tau$, as shown in Fig. \ref{fig:maxwell}, centered at one vertex of the triangle. In this setting, $\xi$ points in the direction normal to the panel, $\eta$ along one edge, and $\tau$ results from the cross product of the corresponding unit vectors ($\mathbf{e}_\tau=\mathbf{e}_\xi\times\mathbf{e}_\eta$). We can then write the normal vector in the integral of Eq. \eqref{eq:maxwell_integral} as  $\mathbf{n}=\mathbf{e}_\xi=(1,0,0)$, and applying the Maxwell tensor to it becomes
\begin{equation}\label{eq:maxwell_normal}
    \mathbf{P}\cdot\mathbf{n} =  \left(\epsilon E_{\xi} E_{\xi}-\frac{1}{2} \epsilon |E|^2 -\frac{1}{2} \epsilon \kappa^{2} \phi^{2} \lambda \right) \mathbf{e}_\xi + \epsilon E_{\xi} E_{\eta} \mathbf{e}_\eta + \epsilon E_{\xi} E_{\tau} \mathbf{e}_\tau
\end{equation}
which is the stress normal to the triangle. Evaluating Eq. \eqref{eq:maxwell_normal} with the unit vectors $\mathbf{e}_\xi$, $\mathbf{e}_\eta$, and $\mathbf{e}_\tau$ expressed in cartesian coordinates recasts the stress in the global frame of reference.

\subsection{Numerical method implementation details} 

\subsubsection{Numerical solution of the boundary integral equation}

We solve Eq. \eqref{eq:bie_limit} numerically on a triangulation of the solvent-excluded surface (SES), using the \texttt{Bempp-cl} library.\cite{betcke2021bempp} \texttt{Bempp-cl} provides high level abstractions of discretized forms of the single and double layer potentials ($V$ and $K$) with an easy Python API, implemented in highly optimized OpenCL code for performance. This allows us to reach large-scale problems on a single workstation.

We assumed a continuous piece-wise linear distribution of $\phi$ and $\partial\phi/\partial\mathbf{n}$ on the triangular panels. In that case, \texttt{Bempp-cl} tracks the values on the vertices of each triangle, rather than the panel itself, and uses a Galerkin approach to arrive at a linear system, such as
\begin{eqnarray}\label{eq:matrix_form}
  \begin{bmatrix}
  1/2 + K_{\Gamma,L}^{\Gamma} & - V_{\Gamma,L}^{\Gamma}  \\
  1/2 - K_{\Gamma,Y}^{\Gamma} &  \dfrac{\epsilon_{1}}{\epsilon_{2}}V_{\Gamma,Y}^{\Gamma} \\
 \end{bmatrix}
 \begin{bmatrix}
  \phi \\
  \dfrac{\partial\phi}{\partial \mathbf{n}}   \\
 \end{bmatrix}
    = 
 \begin{bmatrix}
  \dfrac{1}{\epsilon_{1}} \sum_{k}^{N_{q}} \dfrac{q_{k}}{4\pi |\mathbf{x}_{\Gamma} - \mathbf{x}_{k}|}\\
   0 \\
 \end{bmatrix}
\end{eqnarray}
Then, the solution of this linear system yields the values of $\phi$ and $\partial\phi/\partial\mathbf{n}$ on the vertices, which we used on the discretized form of Eq. \eqref{eq:phi_reac} to obtain $\phi_{reac}$ anywhere in the domain $\Omega_1$. 

Eq. \eqref{eq:matrix_form} is the matrix representation of Eq.\eqref{eq:bie_limit}, which is valid for the single-solute system in Fig. \ref{fig:molecule}. In practice, having just one solute is not an interesting setup to compute forces. The \texttt{BEM} formulation can consider more than one solute by applying the procedure that led to Eq. \eqref{eq:bie_limit} over multiple surfaces,\cite{altman2009accurate,CooperBardhanBarba2014} that can define the molecular surface of another solute or a surface with imposed charge or potential.\cite{CooperClementiBarba2015,CooperBarba2016,cooper2022quantitative}

\subsubsection{The electric field on the molecular surface with a first order boundary element method}

\begin{figure}[H]
    \centering
    \includegraphics[width=0.4\textwidth]{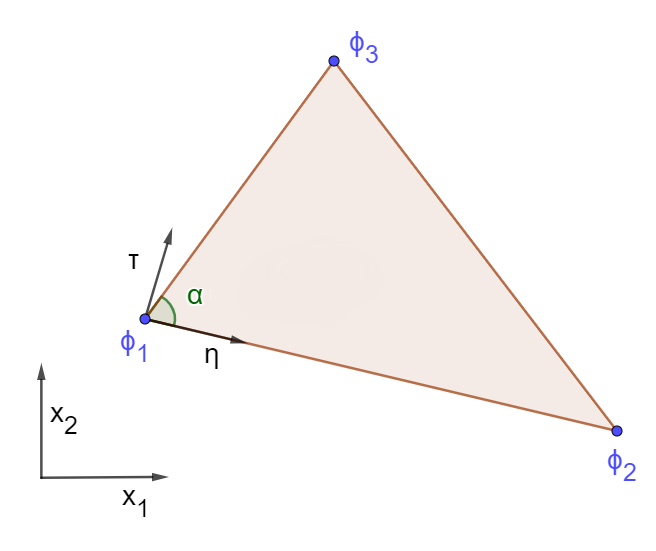}
    \caption{Local coordinate system for a triangular element}
    \label{fig:local_coordinates}
\end{figure}

Solving the system in Eq. \eqref{eq:matrix_form} using continuous piece-wise linear elements with \texttt{Bempp-cl} gives $\phi$ and $\partial\phi/\partial\mathbf{n}$ on the triangle vertices. On the other hand, Eq. \eqref{eq:maxwell_normal} needs the electric field $\mathbf{E} =-\nabla\phi$ in the normal ($E_\xi$) and tangential ($E_\eta$ and $E_\tau$) directions. The normal direction is easy to obtain, as it is an average of $-\partial\phi/\partial\mathbf{n}$ over the vertices of each triangle, however, the tangential directions require some work, and is where the local coordinate system becomes useful. The numerical method assumes a linear distribution of $\phi$ on each panel, which lives on the $(\eta,\tau)$ plane (see Fig. \ref{fig:local_coordinates}), allowing us to write
 \begin{equation}
     \phi(\eta,\tau) = a\eta + b\tau + c.
 \end{equation}
 Using Fig. \ref{fig:local_coordinates}, we can determine $a$, $b$, and $c$ from the values of $\phi$ on the three vertices ($\phi_1$, $\phi_2$, and $\phi_3$), their relative distance ($d_{12}$ and $d_{13}$), and the angle $\alpha$ at vertex 1. The local frame of reference is centered at vertex 1, and $\mathbf{e}_\eta$ points in the direction between vertices 1 and 2. Replacing on vertex 1 gives:
\begin{equation}\label{eq:c}
    \phi(0,0) = a\cdot0 + b\cdot0 + c = \phi_{1}.
\end{equation}
Then, evaluating on $\eta=d_{12}$ gives
\begin{equation}\label{eq:a}
    \begin{aligned}
        \phi(d_{12},0) &= a d_{12} + b \cdot 0 + \phi_{1} = \phi_2 \\
        a &= \frac{\phi_2 - \phi_1}{d_{12}}
    \end{aligned}
\end{equation}
Finally, using the value at vertex 3 ($\phi_3$) gives
\begin{equation}\label{eq:b}
    \begin{aligned}
        \phi(  d_{13}  \cos(\alpha), d_{13} \sin(\alpha)) &= \frac{\phi_2 - \phi_1}{d_{12}}  d_{13} \cos(\alpha) + b   d_{13}  \sin(\alpha) + \phi_1 = \phi_3 \\
    b &= \frac{\phi_3 - \phi_1}{d_{13} \sin(\alpha)} - \frac{\phi_2 - \phi_1}{d_{12} \tan(\alpha) } 
    \end{aligned}
\end{equation}

With the values of $a$, $b$, and $c$ obtained from Eqs. \eqref{eq:a}, \eqref{eq:b}, and \eqref{eq:c}, we can compute the tangential field in each direction analytically as:
\begin{align} \label{eq:fields}
    E_\xi &= -\frac{\partial\phi}{\partial\mathbf{n}} \nonumber\\
    E_\eta &= -\frac{\partial\phi}{\partial\eta} = -\frac{\phi_2 - \phi_1}{d_{12}}\nonumber\\
    E_\tau &= -\frac{\partial\phi}{\partial\tau} = -\frac{\phi_3 - \phi_1}{d_{13} \sin(\alpha)} + \frac{\phi_2 - \phi_1}{d_{12} \tan(\alpha) }
\end{align}

The computation of $\mathbf{E}$ with Eq. \eqref{eq:fields} does not introduce further approximations to the calculation. Then, in the context of a molecular surface represented with flat triangular panels, and a piece-wise linear variation of the potential and its normal derivative, the calculation of the field is exact. This stands out from other implementations of the force calculation with the Poisson-Boltzmann equation \cite{baker2001electrostatics,jurrus2018improvements,AlexovDelphiForce,lu2007} that require numerical approximations on the molecular surface. 

\subsubsection{The energy functional variation approach in boundary integral form}

\paragraph{The charge force ($\mathbf{F}_q$)}
The charge force consists of an integration over the solute volume (see Eq. \eqref{eq:Fq}). Since the charge distribution ($\rho_f$) is a set of Dirac delta functions, the integral becomes a sum over the charges. Like the electrostatic potential leading to Eq. \eqref{eq:phi_reac}, the electric field can also be decomposed into reaction and coulombic components ($\mathbf{E} = \mathbf{E}_{reac} + \mathbf{E}_{coul}$). By the action-reaction principle, two point charges induce equal and opposite forces on them, cancelling out the Coulomb contribution to the total force ($\int_\Omega \rho_f\mathbf{E}_{coul}d\mathbf{x}$=0). Then, we can write
\begin{equation}\label{eq:Fq_reac}
    \begin{aligned}
        \mathbf{F}_{q} &= \int_{\Omega} \rho_{f} \mathbf{E}_{reac} d\mathbf{x}\\
        &= -\sum_{i}^{N}q_{i} \nabla \phi_{reac}(\mathbf{x}_i)
    \end{aligned}
\end{equation}
This could be computed by directly taking the derivative of Eq. \eqref{eq:phi_reac}, however, the gradient of the potential operators $V^\mathbf{x}_\Gamma$ and $K^\mathbf{x}_\Gamma$ are currently not available in \texttt{Bempp-cl}. Then, we calculated $\nabla \phi_{reac}(\mathbf{x}_i)$ by computing $\phi_{reac}$ on near-by locations to each charge, and used a centered difference scheme as
\begin{equation}
    {E}_{i,reac}(\mathbf{x}_{k}) = -\frac{\partial\phi_{reac}}{\partial x_i} \approx - \frac{\phi_{reac}(\mathbf{x}_{k}+h\mathbf{e}_i) - \phi_{reac}(\mathbf{x}_k - h\mathbf{e}_i)  }{2h}
\end{equation}
for $i\in\{1,2,3\}$ the cartesian components and $\mathbf{x}_{k}$ the position of charge $k$. We used $h=0.001$ throughout this study, making sure that the mesh size of this finite difference approximation yielded an error that is low enough to not affect our results.

%

\paragraph{The boundary forces ($\mathbf{F}_{db}$ and $\mathbf{F}_{ib}$)}

The values of $\epsilon$ and $\lambda$ have a sudden jump accross the molecular surface, making the gradients in Eqs. \eqref{eq:Fdb} and \eqref{eq:Fib}  difficult to compute with numerical methods. For example, finite-difference codes like \texttt{APBS},\cite{baker2001electrostatics,jurrus2018improvements} mollify the interface, making $\epsilon$ and $\lambda$ vary across a few mesh points. The boundary integral formulation becomes convenient to avoid these inaccuracies.

Following the work by Cai and co-workers,\cite{cai2012dielectric} we can compute the force across the molecular surface due to the jump in dielectric constant by taking the difference of the terms with $\epsilon$ in the Maxwell stress tensor, evaluated on the inner ($P_{ij}^{in}$) and outer ($P_{ij}^{out}$) sides of $\Gamma$. In the local coordinate system from Eq. \eqref{eq:maxwell_normal}, this gives us the following force density
\begin{align}\label{eq:fdb_density}
    \mathbf{f}_{db} &= \left(\mathbf{P}^{out} - \mathbf{P}^{in} \right)\cdot\mathbf{n} = \left(\mathbf{P}^{out} - \mathbf{P}^{in} \right)\cdot\mathbf{e}_\xi \nonumber\\
    =& \left[ \left( \left(\epsilon E_{\xi} E_{\xi}-\frac{1}{2} \epsilon |E|^2 \right) \mathbf{e}_\xi + \epsilon E_{\xi} E_{\eta} \mathbf{e}_\eta + \epsilon E_{\xi} E_{\tau} \mathbf{e}_\tau \right)^{out} \right. \nonumber\\
    -& \left.\left( \left(\epsilon E_{\xi} E_{\xi}-\frac{1}{2} \epsilon |E|^2 \right) \mathbf{e}_\xi + \epsilon E_{\xi} E_{\eta} \mathbf{e}_\eta + \epsilon E_{\xi} E_{\tau} \mathbf{e}_\tau \right)^{in} \right].
\end{align}
 Considering $\Omega_1$ and $\Omega_2$ the internal and external regions, respectively, we can apply the following interface conditions
\begin{align}\label{eq:interface_conditions}
    \epsilon_1 E_{1,\xi} = \epsilon_2 E_{2,\xi} \nonumber\\
    E_{1,\eta} = E_{2,\eta} \nonumber\\
    E_{1,\tau} = E_{2,\tau}
\end{align}
to cancel out the $\mathbf{e}_\eta$ and $\mathbf{e}_\tau$ components, and write
\begin{align}\label{eq:fdb_density_final}
    \mathbf{f}_{db} &= \left(\left( \epsilon_2 E^2_{2,\xi} -\frac{1}{2} \epsilon |E_2|^2 \right) - \left( \epsilon_1 E^2_{1,\xi}-\frac{1}{2} \epsilon |E_1|^2 \right) \right) \mathbf{e}_\xi\nonumber\\
    &=\frac{1}{2}\left(\epsilon_2(E^2_{2,\xi} - E^2_{2,\eta}-E^2_{2,\tau}) - \epsilon_1(E^2_{1,\xi} - E^2_{1,\eta}-E^2_{1,\tau})\right)\mathbf{e}_\xi\nonumber\\
    &=\frac{1}{2}\left( \epsilon_1 E_{1,\xi} E_{2,\xi} -\epsilon_2 E_{1,\xi} E_{2,\xi} - \epsilon_2(E_{2,\eta}E_{1,\eta} + E_{2,\tau}E_{1,\tau}) + \epsilon_1(E_{2,\eta}E_{1,\eta} + E_{2,\tau}E_{1,\tau}) \right)\mathbf{e}_\xi\nonumber\\
    &=\frac{1}{2}(\epsilon_1-\epsilon_2)\left(E_{2,\xi}E_{1,\xi} + E_{2,\eta}E_{1,\eta} + E_{2,\tau}E_{1,\tau}\right)\mathbf{e}_\xi = -\frac{1}{2}(\epsilon_2-\epsilon_1)(\mathbf{E}_1\cdot\mathbf{E}_2)\mathbf{e}_\xi
\end{align}
Eq. \eqref{eq:fdb_density_final} is in agreement with previous work from Davis and McCammon\cite{Davis1990}. Then, the total force $\mathbf{F}_{db}$ on the molecular surface is
\begin{equation}\label{eq:Fdb_surf}
    \mathbf{F}_{db} = \oint_\Gamma \mathbf{f}_{db} d\mathbf{x} = -\frac{1}{2}(\epsilon_2-\epsilon_1)\oint_\Gamma \mathbf{E}_1\cdot\mathbf{E}_2 \mathbf{e}_\xi d\mathbf{x}
\end{equation}
The electric fields $\mathbf{E}_1$ and $\mathbf{E}_2$ in Eq. \eqref{eq:Fdb_surf} can be computed with Eq. \eqref{eq:fields}. The tangential components of the field are usually much smaller than the normal one,\cite{cai2012dielectric} and $\mathbf{F}_{db}$ can be approximated as\cite{cooper2022quantitative}
\begin{align}\label{eq:Fdb_approx}
    \mathbf{F}^{approx}_{db} = -\frac{1}{2}(\epsilon_2-\epsilon_1)\frac{\epsilon_1}{\epsilon_2}\oint_\Gamma \left(\frac{\partial\phi_1}{\partial\mathbf{n}}\right)^2 \mathbf{n}d\mathbf{x}.
\end{align}
This last expression is very convenient in a boundary integral framework as $\partial\phi/\partial\mathbf{n}$ results directly from solving the system in Eq. \eqref{eq:matrix_form}, without limiting the choice of ansatz to piece-wise linear.

To obtain a surface integral expression for the ionic pressure force ($\mathbf{F}_{ib}$), we can use the same approach that led to Eq. \eqref{eq:Fdb_surf}. This time, we compute the difference of the salt-related terms in the Maxwell stress tensor ($\lambda$ in Eq. \eqref{eq:maxwell_tensor}) on the inner and outer sides of $\Gamma$. This leads to 
\begin{equation}\label{eq:Fib_surf}
    \mathbf{F}_{ib} = -\frac{1}{2} \kappa^{2} \epsilon_{2} \int_{\Gamma} \phi^2 \mathbf{n} d\mathbf{x}
\end{equation}

%% file: results.tex
This section presents force calculations for isolated molecules, and two molecules interacting. We computed the force with the three approaches described in the Methods section, namely, the virtual displacement (Eq. \eqref{eq:virtual_disp}), energy functional (Eqs. \eqref{eq:Fq_reac}, \eqref{eq:Fdb_surf}, and \eqref{eq:Fib_surf}), and Maxwell stress tensor approaches. In the case of the energy functional approach, we also computed the dielectric boundary force with the normal approximation in Eq. \eqref{eq:Fdb_approx} ($\mathbf{F}^{approx}_{db}$). This is summarized in Table \ref{table:methods}, with a naming convention that is used in the rest of this section. To compare, we used the finite difference software \texttt{APBS}~\cite{baker2001electrostatics,jurrus2018improvements}. 

In all cases, the dielectric constant inside the protein was $\varepsilon_1$=4,
and the solvent was set to $\varepsilon_2$=80 and $\kappa$=0.125 \AA$^{-1}$ (corresponding to 150 mM of monovalent ions in the solvent). We used the \texttt{pdb2pqr}\cite{dolinsky2004pdb2pqr} software to parameterize the atomic charge and radii, and then \texttt{Nanoshaper}\cite{decherchi2013general} to generate the surface mesh, unless otherwise noted. Both \texttt{pdb2pqr} and \texttt{Nanoshaper} are called from \texttt{PBJ}. 

The runs were performed on a workstation with two 12-core Intel$^\text{\textregistered}$ Xeon$^\text{\textregistered}$ E5-2680 v3 @ 2.5 GHz CPUs, and 96 GB of RAM.
\begin{table}[]
    \centering
    \begin{tabular}{c|c|c|c}
        \textbf{Name} & \textbf{Description} &\textbf{Eqs.}&\textbf{Refs.}  \\
        \hline
        Method 1 & Virtual displacement & \eqref{eq:virtual_disp} &\cite{Davis1990}\\
        Method 2 & Energy functional variation & \eqref{eq:Fq_reac} \eqref{eq:Fdb_surf} \eqref{eq:Fib_surf} & \cite{Gilson1993} \\
        Method 3 & Approximated energy functional variation &  \eqref{eq:Fq_reac} \eqref{eq:Fdb_approx} \eqref{eq:Fib_surf}& \cite{cooper2022quantitative} \\
        Method 4 & Maxwell stress tensor integration & \eqref{eq:maxwell_integral} \eqref{eq:maxwell_normal} &\cite{Xiao2013}\\
    \end{tabular}
    \caption{Summary and naming convention of force calculation methods with BEM.} 
    \label{table:methods}
\end{table}

\subsection{Results with a single molecule}

As an initial test case, we ran experiments with the different methods detailed in Table \ref{table:methods} on a single lysozyme (PDB code 1lyz), parameterized with the \texttt{AMBER} force field. As the protein is isolated, the total force should be zero, making this a good test case for accuracy. For the same reason, we did not run these experiments with \textit{Method 1}.

Table \ref{tab:1lyz_forces} shows the solvation force and energy for {\it Methods 2, 3,} and {\it 4}, for different surface mesh refinements. As expected, all methods are converging to zero as the mesh density increases, however, {\it Method 4} generates the most accurate results, and {\it Method 3} the least. This is an expected result for two reasons. First, {\it Method 3} behaves worse because it uses an approximation on the dielectric boundary force (Eq. \eqref{eq:Fdb_approx}) that neglects the electric field in off-normal directions. Second, {\it Method 2} involves the sum of two large and opposite components, namely, $\mathbf{F}_q$ and $\mathbf{F}_{db}$ (see Table \ref{tab:1lyz_decomp} for their magnitude). This is a difficult situation for the numerical method, as small errors in $\mathbf{F}_q$ and $\mathbf{F}_{db}$ may result in a large error in their difference. This does not happen with {\it Method 4}. The force calculations with \texttt{APBS} in Table \ref{tab:1lyz_forces_APBS} also use the energy functional approach (similar to {\it Method 2}), and hence, they have the same accuracy issues. Even though the solution with \texttt{APBS} seems to be converging to zero, it performs worse than {\it Method 2} and {\it Method 3}.

To analyze the convergence, we can use the concept of {\it observed order of convergence} ($p$)\cite{roache1998verification,CooperBardhanBarba2014}
\begin{equation}\label{eq:p}
    p = \frac{\log\left(\frac{f_1-f_2}{f_2-f_3}\right)}{\log(r)}
\end{equation}
where $f_1$, $f_2$, and $f_3$ are the solutions with a coarse, medium, and fine mesh, respectively, and $r$ is the mesh density ratio between them. If the details of the solution are appropriately resolved, $p$ should match the order con convergence of the numerical method and we say it is in the {\it asymptotic convergent region}. Our boundary integral method uses linear elements that give first order convergence. Considering the mesh densities 4, 8, and 16 vertices per \AA$^2$ from Table \ref{tab:1lyz_forces} in Eq. \eqref{eq:p}, we get $p$=1.2 for {\it Method 4} and $p$=1.4 for {\it Method 2} and {\it Method 3}, which indicates that they all are asymptotically converging. Using the three finest meshes of APBS in Table \ref{tab:1lyz_forces_APBS} results in $p$=1.48, which is similar to our BEM approach, however, the results are still far from the real solution ($|\mathbf{F}|$=0). It is important to consider that force calculations with \texttt{APBS} use a 4$^{th}$-order spline to mollify the dielectric interface and compute the electric field on the molecular surface, adding an extra layer of approximations. 

In the work by Sørensen {\it et al.},\cite{sorensen2015comprehensive} the authors performed a careful analysis of the impact of mesh spacing on solvation and binding free energies for various finite difference codes (\texttt{APBS} among them). They recommended a spacing of $\Delta x$=0.5 or less for acceptable binding energy results. On the other hand, a similar analysis with BEM\cite{CooperBardhanBarba2014} concludes that a mesh with 2 vertices/\AA$^2$ is the coarsest refinement that yields acceptable results for solvation and binding energies. Table \ref{tab:1lyz_forces_APBS} shows that a mesh spacing of $\Delta x$=0.117, which is 4$\times$ finer than Sørensen {\it et al.}'s recommendation, is less accurate than using 2 vertices/\AA$^2$ with {\it Method 4}, and 8 vertices/\AA$^2$ with {\it Method 2}. This indicates that a BEM approach the same mesh that is valid for solvation energy calculations is useful to compute the force. This is not the case in finite differences, which has been reported in the past \cite{boschitsch2015adaptive}.

\begin{table}[H]
  \begin{center}
    \caption{Solvation energy (kcal/mol) and force magnitude (kcal/mol\AA) for 1lyz, mesh density in vertices/\AA$^2$.}
    \label{tab:1lyz_forces}
    \begin{tabular}{c|c|c|c|c} 
      \textbf{Mesh dens.} & \textbf{Method 2} & \textbf{Method 3} & \textbf{Method 4} & \textbf{$\Delta G_{solv}$}\\
      \hline
      2 & 5.2553 & 7.0078 & 0.6234 & -484.70  \\ 
      4 & 2.0308 & 3.3422 &  0.2325 & -465.74  \\ 
      8 &  0.8131 & 1.9724 &0.1013 &  -458.31  \\ 
      16 & 0.3649 & 1.4639 & 0.0458 & -455.22  
    \end{tabular}
  \end{center}
\end{table}

\begin{table}[H]
  \begin{center}
    \caption{Force decomposition (magnitude in kcal/mol\AA) for 1lyz using {\it Method 2} and {\it Method 3}. Mesh density in vertices/\AA$^2$.}
    \label{tab:1lyz_decomp}
    \begin{tabular}{c|c|c|c|c|c|c} 
      \textbf{Mesh} & \multicolumn{3}{|c|}{\bf Method 2} & \multicolumn{3}{|c}{\bf Method 3} \\
      \textbf{dens.}& $|\mathbf{F}_{q}|$ & $|\mathbf{F}_{db}|$ & $|\mathbf{F}_{ib}|$ & $|\mathbf{F}_{q}|$ & $|\mathbf{F}_{db}|$ & $|\mathbf{F}_{ib}|$ \\
      \hline
        2 & 38.0419 & 32.6898 & 0.1419 & 38.0419 & 30.9821 & 0.1419  \\ 
        4 & 29.2129 & 27.0556 & 0.1405 & 29.2129 & 25.7790 & 0.1405  \\ 
        8 & 27.6445 & 26.6999 & 0.1411 & 27.6445 & 25.5650 & 0.1411  \\ 
        16 & 26.2625 & 25.7651 & 0.1410 & 26.2625 & 24.6971 & 0.1410  
    \end{tabular}
  \end{center}
\end{table}

\begin{table}[H]
  \begin{center}
    \caption{APBS force magnitude (kcal/mol\AA) for 1lyz, mesh density in $\Delta x$ \AA, box size 60$\times$60$\times$60.}
    \label{tab:1lyz_forces_APBS}
    \begin{tabular}{c|c|c} 
      \textbf{$\Delta x$} & \textbf{Nodes} & |\textbf{F}| \\
      \hline
        0.938 & 65$\times$65$\times$65 & 73.439 \\
        0.469 & 161$\times$161$\times$161 & 64.234 \\
        0.208 & 321$\times$321$\times$321 & 17.963  \\
        0.117 & 513$\times$513$\times$513 & 1.4699
    \end{tabular}
  \end{center}
\end{table}

\subsection{Results for two spherical molecules}

Force calculations are useful to study the interaction between two or more molecules. As a simple model problem, we computed the force induced by a spherical molecule on another spherical molecule ($\mathbf{F}_{bind}$). In general, $\mathbf{F}_{bind}$ is the difference in force between an interacting state, where spheres are close-by, and a non-interacting one. As there are only two spheres, the molecules are isolated in the non-interacting state, and the force is zero. For that reason, we only need to compute the force in the interacting state.

Both spheres had a centered charge of 2$q_e$ and a radius of 1 \AA, and we generated the meshes with \texttt{MSMS}~\cite{Sanner96}. In this case it makes sense to use {\it Method 1} because the free energy depends on the relative distance between the spheres, which changes in the virtual displacement calculations (offset by $h\mathbf{e}_i$ with $h=0.001$ \AA) of Eq. \eqref{eq:virtual_disp}.

\begin{figure}[h]
    \centering
    \includegraphics[width=0.75\textwidth]{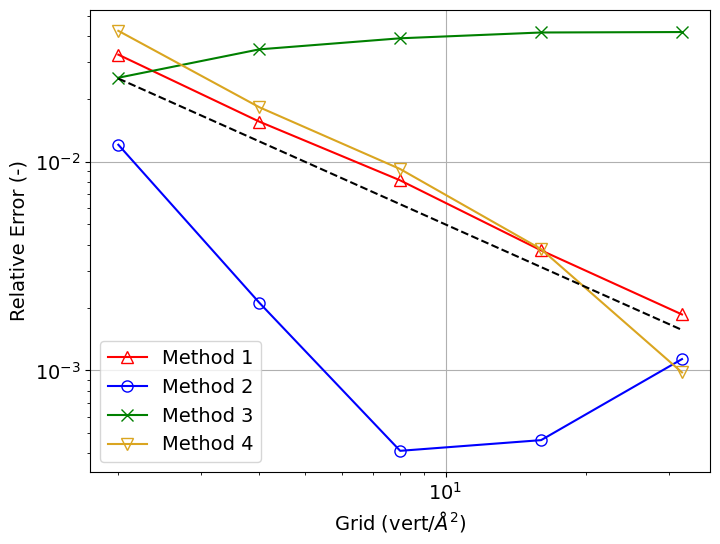}
    \caption{Error between methods for two 1 \AA~spheres with a centered 2$q_e$ charge at 3 \AA~ center-to-center distance. Dotted line indicates linear convergence.}
    \label{fig:force_spheres_loglog}
\end{figure}

Table \ref{tab:sphere_forces} shows a mesh refinement study of the force and binding energy when the spheres are 3 \AA~ away, where $\Delta G_{bind}$ is the energetic difference between interacting and isolated states. As a reference solution, we used closed expressions for the solvation energy of two spheres,\cite{lotan2006analytical,siryk2021charged} and computed the force by applying them to the virtual displacement approach in Eq. \eqref{eq:virtual_disp}. This reference value was $\mathbf{F}_{ref}$=1.9425 kcal/mol\AA, which is the base in the error plots of Fig. \ref{fig:force_spheres_loglog}. It is interesting to note that even though {\it Method 2} is more accurate than {\it Method 4}, the latter is converging with the expected first order trend (as also {\it Method 1}), when {\it Method 2} is not. Similarly to the isolated case with lysozyme, it is difficult to obtain the right convergence with {\it Method 2}, as it involves the subtraction of two large numbers ($\mathbf{F}_q$ and $\mathbf{F}_{db}$). This makes {\it Method 4} a more robust option.

\begin{figure}[h]
    \centering
    \includegraphics[width=0.8\textwidth]{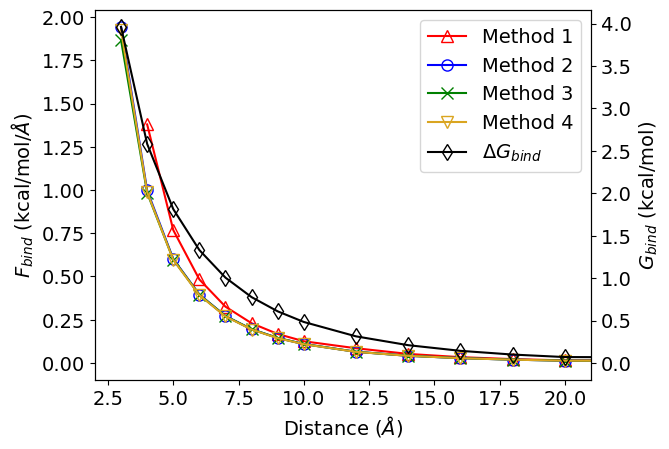}
    \caption{Induced force and binding energy between two 1 \AA~spheres with a centered 2$q_e$ charge, at different center-to-center distances.}
    \label{fig:force_spheres_dist}
\end{figure}

Fig. \ref{fig:force_spheres_dist} shows the induced force at different center-to-center distances for the same two spheres, using a 8 vertices/\AA$^2$ mesh and $h=1$ \AA\ for {\it Method 1}. Even though the errors in Fig. \ref{fig:force_spheres_loglog} are different between methods 2, 3, and 4, in the context of Fig. \ref{fig:force_spheres_dist} these curves are overlapped. In this case, {\it Method 1} struggles as the spheres get closer because $\Delta G_{bind}$ (and hence, $\Delta G_{solv}$) grows, then, small errors in $\Delta G_{solv}$ generate large errors in the force calculated with Eq. \eqref{eq:virtual_disp}. Also to get a accurate gradient is necessary to get more points on the highest variations of $\Delta G_{bind}$ which in this case implies the use of a variable spacing $h$

\begin{table}[h]
  \begin{center}
    \caption{Solvation force x-component (kcal/mol\AA) for sphere 2 charge 2q with 3 \AA~  between centers, mesh density in vertices/\AA$^2$. Using the virtual work approach with an analytical solution for the energy gives a force of 1.9425 kcal/mol\AA.}
    \label{tab:sphere_forces}
    \begin{tabular}{c|c|c|c|c|c} 
      \textbf{Mesh} &  & & & & $\Delta G_{bind}$ \\
    \textbf{dens.} & \textbf{Method 1} & \textbf{Method 2} &\textbf{Method 3} & \textbf{Method 4} & kcal/mol \\
      \hline
        2 & 1.8794 & 1.9192 & 1.8936 & 1.8604  &3.9345\\ 
        4 & 1.9124 & 1.9385 & 1.8756 & 1.9072 & 3.9523\\ 
        8 & 1.9268 & 1.9434 & 1.8669 & 1.9247 & 3.9612\\ 
        16 & 1.9353 & 1.9435 & 1.8619 & 1.9352 & 3.9667\\ 
        32 & 1.9390 & 1.9438 & 1.8615 & 1.9407 & 3.9691 
    \end{tabular}
  \end{center}
\end{table}

\subsection{Results for the barnase-barstar complex}

\begin{figure}
    \centering
    \includegraphics[width=0.8\textwidth]{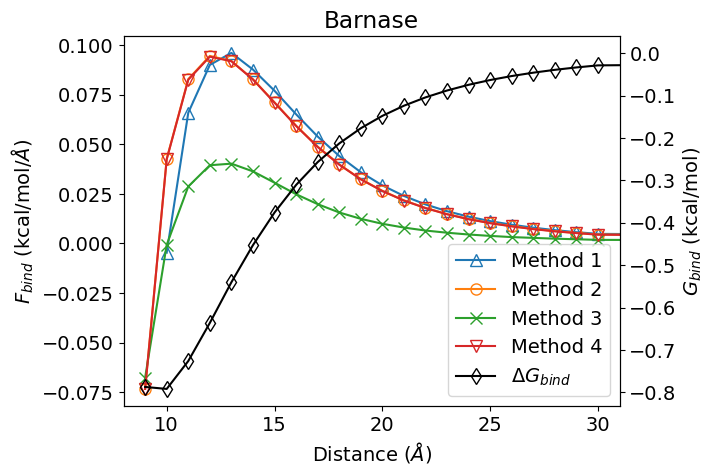}
    \caption{Z-component of force induced by barstar on barnase and binding energy, at different offsets of barstar in the z axis with respect to its original position from the PDB crystal structure.}
    \label{fig:force_barnase_dist}
\end{figure}

\begin{figure}
    \centering
    \includegraphics[width=0.8\textwidth]{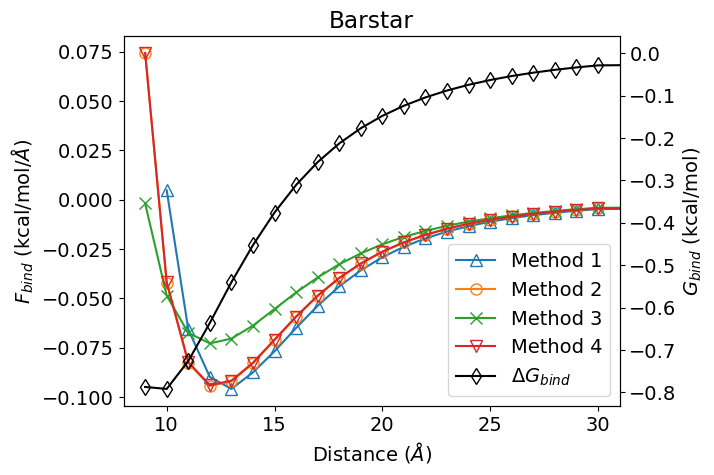}
    \caption{Z-component of force induced by barnase on barstar and binding energy, at different offsets of barstar in the z axis with respect to its original position from the PDB crystal structure.}
    \label{fig:force_barstar_dist}
\end{figure}

The barnase-barstar complex is a standard case study for binding energy calculations.\cite{ bertonati2007poisson,AlexovDelphiForce,nguyen2017accurate} Here, we used chains B (barnase) and E (barstar) of the structure under the PDB ID 1brs\cite{buckle1994protein}, and moved barstar up in the z direction, away from barnase. In the closest position, barstar was displaced 9 \AA~ in the $z$ direction (see Figs. \ref{fig:force_barnase_dist} and \ref{fig:force_barstar_dist}), which was the smallest displacement that did not generate clashes between the two molecular surfaces. We meshed the solvent excluded surface of both molecules with 8 vertices/\AA$^2$ and use $h=1$ \AA~ for {\it Method 1}.

Similar to the sphere case in Fig. \ref{fig:force_spheres_dist}, the non-interacting state has both molecules isolated, where the force should be exactly zero, making the total force equal to $\mathbf{F}_{bind}$. However, from Table \ref{tab:1lyz_forces} we see that there is a numerical error, which decreases as the mesh is refined. To substract out this error, we explicitly computed the force placing barstar and barnase far away (at 100 \AA), and subtracted that out from the calculations performed at each distance. 

Figs. \ref{fig:force_barnase_dist} and \ref{fig:force_barstar_dist} show the z-component of $\mathbf{F}_{bind}$ and $\Delta G_{bind}$ of barnase and barstar, respectively, as a function of the distance barstar was moved from its original position in the PDB structure. We can see that {\it Method 2} and {\it Method 4} are overlapping, whereas {\it Method 3} performs worse. Even though for large distances the accuracy of {\it Method 3} seems acceptable, as barnase and barstar get closer, the off-normal components of the field become more important, and the approximation in Eq. \eqref{eq:Fdb_approx} is inadequate. Results with {\it Method 1} are close to {\it Methods 2} and {\it 4}. Computing the force with {\it Method 1} for small distances is challenging because we need to avoid mesh clashing in the virtual displacements calculations of Eq. \eqref{eq:virtual_disp}. Moreover, when both molecules are close, $\Delta G_{bind}$ changes only slightly (see black curve for distances close to 10 \AA~ in Figs. \ref{fig:force_barnase_dist} and \ref{fig:force_barstar_dist}), making it difficult to capture with the numerical derivative of Eq. \eqref{eq:virtual_disp}. At large distances, all methods seem to be performing similarly.   

In our setup, barstar is placed above barnase in the z-axis. Then, a positive z-component of $\mathbf{F}_{bind}$ in Fig. \ref{fig:force_barnase_dist} indicates an attractive interaction, whereas attraction happens when the force is negative in Fig. \ref{fig:force_barstar_dist}. As barstar approaches barnase the interaction is initially attractive, and then flips to repulsive. This is an indication that at small distances we would see a deceleration of the approaching molecules, in what is known as {\it soft landing}.\cite{Shashikala2019}

%% file: conclusions.tex
The Poisson-Boltzmann equation is usually restricted to electrostatic potential and free energy calculations, however, the force provides useful insights, for example, to study molecular interaction and binding, which can be tested experimentally~\cite{hernando2015quantitative}. As the force is a derivative of the energy, it is a challenging quantity to calculate numerically. Starting from piece-wise linear boundary elements, our approach computes the electric field on the molecular surface exactly, without adding numerical approximations to the standard Poisson-Boltzmann calculation of the potential. Here, we presented a thorough analysis of different formulations to obtain the force with a boundary element method. Where we compared four different methods, and found that the most accurate one is based on the Maxwell stress tensor, followed by a method that relies on the variation of the energy functional. We also introduced an approximation to the energy functional approach that considers the normal component of the electric field only. This method gave acceptable results when the molecules were far apart. We verified our approach against known solutions for single molecules and two interacting spheres. We also compared the accuracy with the finite difference code, and saw that the boundary integral approach outperforms the finite difference method for equivalent meshes.

In the future, we plan to use this efficient approach in applications where high accuracy is required for reliable simulations. Some examples are the force induced on large structures, such as viruses-materials~\cite{cooper2022quantitative}, and adsorption calculations~\cite{GuzmanJPCB2022}, where we need to detect the influence of small changes in orientation~\cite{CooperClementiBarba2015,tsori2020bistable,urzua2022predicting}.